\documentclass[
aps,%
11pt,%
final,%
notitlepage,%
oneside,%
onecolumn,%
nobibnotes,%
nofootinbib,
superscriptaddress,%
noshowpacs,%
centertags]%
{revtex4}

\RequirePackage{graphicx}

\def\refitem#1{\relax}

\begin{document}

\title{Simple Solution to the Strangeness Horn Description  Puzzle 
}

\author{\firstname{K.A.} \surname{Bugaev}}
\email{Bugaev@th.physik.uni-frankfurt.de}
\affiliation{Bogolyubov Institute for Theoretical Physics, Metrologichna str. 14$^B$, Kiev 03680, Ukraine}

\author{\firstname{D.R.} \surname{Oliinychenko}}
\email{Dimafopf@gmail.com}
\affiliation{Bogoliubov Laboratory of Theoretical Physics, JINR, Joliot-Curie Str.  6,   Dubna 141980, Russia}

\author{A.S. Sorin}
\email{Sorin@theor.jinr.ru}
\affiliation{Bogoliubov Laboratory of Theoretical Physics, JINR, Joliot-Curie Str.  6,   Dubna 141980, Russia}

\author{G.M. Zinovjev$^1$}
\email{Gennady.Zinovjev@cern.ch}

\begin{abstract}
We propose to use the thermal model with the multi-component hard-core radii to describe the hadron yield ratios 
from the low AGS to the highest RHIC energies. It is demonstrated  that the variation of the hard-core radii of pions and 
 kaons enable us to drastically improve the fit quality of the measured  mid-rapidity data  and for the first time to completely   describe  the Strangeness Horn behavior as the function of the energy of collision without spoiling the fit quality of other ratios. The   best global fit is found for the vanishing hard-core radius of pions and for the hard-core radius of  kaons being equal to 0.35 fm, whereas the hard-core radius of all other mesons is fixed to 0.3 fm and that one of baryons is fixed to  0.5 fm. 
It is argued that the multi-component hadron resonance gas model opens us a principal possibility  to determine the second virial coefficients of hadron-hadron interaction.
\end{abstract}

\maketitle

\section{Introduction} 

The hadron resonance gas model \footnote{We apologize for not quoting even  the major  works on this model which are well known, but the list is so long that we have to choose just the  papers strictly related to our discussion. } \cite{KABAndronic:05,KABAndronic:12} is the only  theoretical tool allowing us to extract information about the chemical freeze-out (FO) stage of the relativistic heavy ion collisions. 
Although its systematic application  to the experimental data description  began about fifteen years ago \cite{KABbig_radii}, many features of this model are not well  studied 
\cite{KAB_two_comp_VdW,KABOliinychenko:12}.  Thus, very recently in a critical analysis of the 
hadron resonance gas model \cite{KABOliinychenko:12} it was shown that for the description of the hadron multiplicities the baryon charge conservation and the isospin conservation, used in one of the most 
successful versions of this model \cite{KABAndronic:05}, should be essentially modified, whereas for 
the description of the hadron yield ratios these conservation laws are not necessary at all.
Although   the discussion  about the reliable chemical FO criterion has a long history \cite{KABAndronic:05,KABCleymansFO}, only recently it was demonstrated that none of the previously suggested chemical FO criteria, including the most popular one of constant energy per particle $E/N \simeq 1.1$ GeV  \cite{KABCleymansFO}, is  robust \cite{KABOliinychenko:12},  if the realistic particle table with the hadron masses up to 2.5 GeV is used.  At the same time in \cite{KABOliinychenko:12} 
 it was  shown that despite an essential difference with the approach used in \cite{KABAndronic:05},
 the both versions of the hadron resonance gas model demonstrate almost the same value 7.18  for   the  entropy per particle  at chemical FO.  Thus, it turns out that the criterion of the constant entropy per particle at chemical FO is, indeed, a  reliable one. 
It is interesting that the constant entropy per particle at chemical FO was also found in \cite{KABTiwari:12},
but the way of hard core repulsion  used in this work is too different from the traditional  one used in the hadron resonance gas model \cite{KABAndronic:05, KABOliinychenko:12, KABAndronic:12} and,  hence, in contrast to the results of  \cite{KABAndronic:05, KABOliinychenko:12,KABAndronic:12}, the model used in \cite{KABTiwari:12} leads to a  simultaneous fulfillment of a few chemical FO criteria. 

One of the traditional difficulties  of the hadron resonance gas model is related to the  Strangeness Horn description  which up to now is far from being satisfactory, although very different formulations of  the thermal model are used for this purpose  \cite{KABAndronic:05,KABOliinychenko:12, KABTiwari:12,KABStrangeness_horn_sigma_meson}.
Note that the principal importance   to improve the Strangeness Horn description can be easily understood from the fact that just the non-monotonic behavior of the $K^+$/$\pi^+$ ratio as the function of the center of mass energy of collision is often claimed to be one of a few existing signals of the onset on deconfinement \cite{Gazd_Horn:99, Step, Gazd_rev:10}. 
The previous attempts  \cite{KABAndronic:05, KABTiwari:12,KABStrangeness_horn_sigma_meson}
to describe  the Strangeness Horn behavior without spoiling the quality of  other particle ratios fit  and the 
  thorough  analysis performed in \cite{KABOliinychenko:12}   led us to a conclusion that  further improvement  of the hadron resonance gas model can be achieved, if  we consider the pion and kaon hard-core radii, as an independent fitting parameters. Evidently,  this would allow us  to have two additional fitting parameters  to improve the fit quality of the Strangeness Horn without spoiling the other hadron yield  ratios. The physical idea behind such an approach is that the
hadronic hard-core radii  are the effective parameters which include the contributions of the repulsion and attraction. Since the parameters of hadron-hadron interaction are, generally speaking,  individual for each kind of hadrons, then each kind of hadrons can have  its own
hard-core radius. 

The work is organized as follows. Section II contains the main equations of the model. The results are discussed  in Section III, while the last Section contains our conclusions.

\section{Multi-component hadron gas and hard-core radii}

The hadron resonance gas model is a successful compromise between the right choice of  the physically relevant degrees of freedom and the simple parameterization of their interaction. Its theoretical justification is based on a simple fact found rather long ago \cite{KABRaju:92} that for temperatures below 170 MeV the interacting mixture of   stable  hadrons and their   resonances  behaves  as the mixture of  nearly  ideal gases of stable  particles which in this case include both the  hadrons and the resonances taken with their averaged masses.   The reason for such a behavior is nearly a complete cancellation between the attraction and repulsion contributions.  The resulting deviation from the ideal gas (a weak repulsion)   is usually attributed to the second virial coefficients $b_{ij}$ defined for the hadrons of $i$-th and $j$-th kinds. 
Since the equations of state of the hadron resonance  gas have the Van der Waals type repulsion, the coefficients $b_{ij}$ are called as  excluded volumes.
Up to now the  hadron resonance  gas model employed only two basic parameters related to hadron-hadron repulsion: the model of   
\cite{KABbig_radii} had one  excluded volume for pions and another for all other hadrons, whereas 
 the model \cite{KABPBM:99} suggested  to consider one common hard-core radius for mesons $R_m$ and another  hard-core radius for baryons $R_b$. However, none of the models  developed in  \cite{KABbig_radii,KABPBM:99} were correct \cite{KAB_two_comp_VdW}, since they did not include the second virial coefficient of 
 the crossed type $b_{ij}$ with $i \neq j$, i.e. between the hadrons of different kinds.
On the one hand  the realization of this fact led to a systematic description of the data with a single 
hard-core radius for all hadrons \cite{KABAndronic:05}, and on the other hand it also led to the development  
of  the multi-component Van der Waals gas models  \cite{KABKostyuk:99,KAB_two_comp_VdW,Bugaev_Lorentz_cont_2}.

The success of the hadron resonance gas model, the one \cite{KABAndronic:05} or two component \cite{KAB_two_comp_VdW,KABOliinychenko:12},  in the data description  may look surprising at the first glance, but this is not just a single example of  a simple statistical model that is able to efficiently account for the  complex features of interaction between the constituents. One should remember,  although  the interaction between the  clusters of many molecules in the real gases or interaction  between the nuclear fragments is no less, but more complex than interaction of hadrons, the successful statistical models for such systems are well known 
\cite{KABFisher_67,KABDillmann:91, KABLFK:94,KABBondorf,KABBugaev:00}, and, nevertheless, these models are 
able not only to  describe the   low density states of  real   gases \cite{KABFisher_67,KABDillmann:91, KABLFK:94} or  that ones of  nuclear fragments   \cite{KABBondorf,KABBugaev:00}, but they are able to successfully model  the condensation of these gases into the corresponding liquids at rather high densities.
And one should also  remember that these models  employ  a few statistical parameters only and  use  rather simple, but physically adequate (!)
parameterization for the many-body effects. Thus,  the whole point is that in all these successful examples \cite{KABFisher_67,KABDillmann:91, KABLFK:94,KABBondorf}
the employed parameters which  characterize the interaction  are effective from the very beginning  and  
only at very low densities the models recover   the virial expansion up to the  second virial coefficients. 
Therefore, one should not be surprised  that the second virial coefficients $b_{ij}$ of the hadron resonance gas 
are some effective parameters which account  for (to large extent) a cancellation of the attractive and repulsive contributions, and which, in principle,  could be  individual characteristics for  each hadronic pair. However, the overall success of the hadron resonance gas model evidences that the number of independent parameters  should be essentially smaller then the  number of hadron types.  Thus, below we  demonstrate that the available data favor 
just the set of the excluded volumes defined as $b_{ij} \equiv \frac{2 \pi}{3} (R_i+R_j)^3$ via the 
hard-core radii of pions $R_\pi$, kaons $R_K$, baryons $R_b$ and  the radius for all other mesons $R_m$. In what follows  we give the main equations of the multi-component formulation referring  to \cite{KAB_two_comp_VdW,KABOliinychenko:12}
for a detailed derivation.

Consider  the Boltzmann gas of $N$ hadron species in a volume $V$ that has  the temperature $T$, the baryonic chemical potential $\mu_B$, the  strange chemical potential $\mu_S$ and the chemical potential of the isospin third component $\mu_{I3}$. The system  pressure $p$ and the $K$-th charge density $n^K_i$ ($K\in\{B,S, I3\}$) of the 
i-th hadron sort are given by the expressions  ($\cal B$ denotes a symmetric  matrix of the second  virial coefficients with the elements $b_{ij}$)
\begin{eqnarray}\label{EqI}
p = T \, \sum_{i=1}^N \xi_i \,
,   \quad n^K_i = Q_i^K{\xi_i} {\textstyle \left[ 1+\frac{\xi^T {\cal B}\xi}{\sum\limits_{j=1}^N \xi_j} \right]^{-1}} \,,  \quad  \xi=\left(
\begin{array}{c}
 \xi_1 \\
 \xi_2 \\
... \\
\xi_s
\end{array}
\right)\,, 
\end{eqnarray}
where the variables $\xi_i$ are the solution of the following system
\begin{eqnarray}\label{EqII}
 \xi_i=\phi_i (T)\,   \exp\left(\frac{\mu_i}{T} -\sum\limits_{j=1}^N 2\xi_j b_{ij}+\frac{\xi^T{\cal B}\xi}{\sum\limits_{j=1}^N\xi_j}\right) \,, \quad   \phi_i (T) = \frac{g_i}{(2\pi)^3}\int \exp\left(-\frac{\sqrt{k^2+m_i^2}}{T} \right)d^3k  \,.
\end{eqnarray}
Here the full chemical potential of the $i$-th hadron sort $\mu_i \equiv Q_i^B \mu_B + Q_i^S \mu_S + Q_i^{I3} \mu_{I3}$ is expressed in terms of the corresponding charges $Q_i^K$  and their  chemical potentials,  $ \phi_i (T) $ denotes 
the thermal particle  density of  the $i$-th hadron sort of mass $m_i$ and degeneracy $g_i$, and  $\xi^T$  denotes  the row of  variables $\xi_i$. 

In a special case  when all  the elements of the second  virial coefficients  matrix are equal  $b_{ij}=v_0$  Eqs. (\ref{EqI})--(\ref{EqII}), evidently,   reproduce the one component model with the pressure 
\begin{eqnarray}\label{EqIII}
p  &=&  T \sum\limits_{i=1}^N \, \phi_i (T)\,    \exp \left[  \frac{\mu_i - p\,v_0}{T} \right] \,,
\end{eqnarray}
which defines the particle density of $i$-th kind of hadron  as $n_i = \frac{\phi_i (T)}{1+p\,v_0/T} \exp \left[  \frac{\mu_i - p\,v_0}{T} \right] $. The latter shows that 
 the ratios of  two particle densities  defined by  (\ref{EqIII})
match that ones of the mixture of the corresponding  ideal gases  for an arbitrary
value of $v_0$, while the particle densities themselves may essentially differ
from the particle densities of the ideal gas.

It is known that the resonance width is important at low temperatures \cite{KABAndronic:05}. Similarly to \cite{KABAndronic:05},  the width $\Gamma_i$ of the resonance of mean  mass $m_i$  is   modeled  by replacing the Boltzmann
distribution function in the particle  thermal density (\ref{EqII})  by its average over the
 Breit-Wigner mass distribution as
\begin{eqnarray}\label{EqIX}
&&\int \exp\left(-\frac{\sqrt{k^2+m_i^2}}{T} \right)d^3k \rightarrow 
 \frac{\int_{M_0}^{\infty} \frac{dx}{(x-m_i)^2+\Gamma_i^2/4}\int \exp\left(- \frac{\sqrt{k^2+x^2}}{T} \right)d^3k}{\int_{M_0}^{\infty} \frac{dx}{(x-m_i)^2+\Gamma_i^2/4}} \,,
\end{eqnarray}
where $M_0$ is the  dominant decay channel mass. Such a substitution provides a simple, but reliable approximation to account for the resonance width. 

The contribution of the resonance decays is  accounted for  as usual: the total 
density of hadron $X$  consists of the thermal part  $n^{th}_X$  and the decay ones:
\begin{eqnarray}\label{EqVIII}
n^{tot}_X = n^{th}_X+ n^{decay} = n^{th}_X + \sum_{Y} n^{th}_Y \, Br(Y \to X) \,,
\end{eqnarray}
where $Br(Y \to X)$ is the decay branching of  the Y-th hadron  into the hadron X. The masses, the  widths and the strong decay branchings of all hadrons  were  taken from the particle tables  used  by  the  thermodynamic code THERMUS \cite{THERMUS}.

The strange charge conservation  completes the list of equations used. Since in strong decays 
the strangeness is conserved, then it is sufficient to impose the vanishing of the total strangeness 
for thermal densities at chemical FO, i.e. to determine the strange chemical potential $\mu_S$ from the 
equation $\sum\limits_{i=1}^N  n_i^S = 0$.
As it was shown  recently \cite{KABOliinychenko:12}
the baryonic charge and isospin conservation laws should not be imposed to fit the hadron multipllicities
since they lead to unphysically huge FO volumes.  Therefore, in this work for the data at given energy of collision  we use the following fitting 
parameters: temperature $T$, baryonic chemical potential $\mu_B$ and the chemical potential of the third projection of  isospin  $\mu_{I3}$. Note that such a procedure is completely consistent with fitting the hadron multiplicities instead of  hadron yield ratios \cite{KABOliinychenko:12}, the main difference is only that to fit  the hadron multiplicities one has to use  the chemical FO volume $V$ as  an additional parameter.  As it was explained earlier the global fitting parameters are the  hard-core radii of pions $R_\pi$, kaons $R_K$, baryons $R_b$ and  that one  for all other mesons $R_m$.
In addition, to demonstrate the pure effect of the radii variation we do not include  any strangeness suppression factor into  simulations. Then one should expect  some minor problems with the description of multi-strange baryons.

\section{Results}

The recent  comprehensive analysis \cite{KABOliinychenko:12} performed for  different hard-core radii of  all mesons $R_m$
and all baryons $R_b$ clearly showed us that the good description of the data can be achieved for many pairs of  these radii. However, the ratios are more stable during the fitting, if  $R_m = 0.3$ fm and $R_b = 0.5$ fm.
Right these values of hard-core radii were fixed and then we fitted the data by the $\chi^2/dof$-criterion for  different values of  the pion $R_\pi$ and kaon $R_K$ hard-core radii taken below 0.5 fm each.
The minimal value of  $\chi^2/dof \simeq 1.018$ for the  energies in the range   $\sqrt{s_{NN}} =  2.7, 3.3, 3.8, 4.3, 4.9, 6.3, 7.6, 8.8, 12, 17, 130, 200$ GeV (for details see below) was obtained for $R_\pi = 0$ fm and $R_K = 0.35$ fm.

Since in the present  approach  there is no principle difference between fitting the absolute hadron  yields at mid-rapidity  or their ratios, we prefer to fit ratios in order to reduce the volume of  numerical efforts. 
In our choice of the data sets we basically  followed Ref. \cite{KABAndronic:05}. Thus, at  the AGS energy range of collisions ($\sqrt{s_{NN}} = 2.7 -4.9$ GeV) the data are  available for the kinetic beam energies from 2 to 10.7 AGeV.  For the beam energies 2, 4, 6 and 8 AGeV there are only a few data points available: the yields for pions \cite{AGS_pi1, AGS_pi2}, for protons \cite{AGS_p1, AGS_p2}, for kaons  \cite{AGS_pi2} (except for 2 AGeV), for  $\Lambda$ hyperons the integrated over $4 \pi$ data are available \cite{AGS_L}. For the beam  energy 6 AGeV there exist  the $\Xi^-$ hyperon  data integrated over $4 \pi$ geometry   \cite{AGS_Kas}. However, the data for  the  $\Lambda$ and $\Xi^-$ hyperons
have to be corrected \cite{KABAndronic:05}, and instead of the raw experimental data we used their  corrected values of  Ref. \cite{KABAndronic:05}.
For the highest AGS center of mass energy $\sqrt{s_{NN}} = 4.9$ GeV (or the beam energy 10.7 AGeV) in addition
to the mentioned data for pions, (anti)protons and  kaons  there exist data for $\phi$ meson \cite{AGS_phi},
for  $\Lambda$ hyperon \cite{AGS_L2} and  $\bar \Lambda$ hyperon \cite{AGS_L3}.

\begin{figure}[htbp]
\centerline{  \includegraphics[height=7 cm]{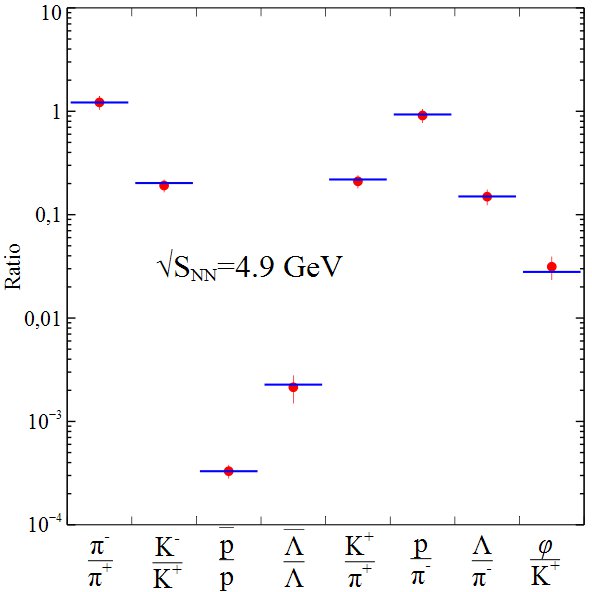} 
\hspace*{.5cm}
\includegraphics[height=7 cm]{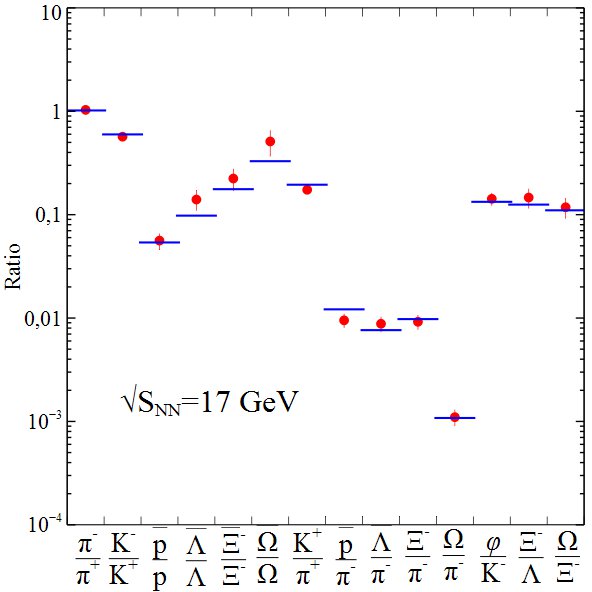} 
}  
 \caption{The particle  yield  ratios described   by  the present multi-component hadron gas model. The best fit for   $\sqrt{s_{NN}} = 4.9$ GeV is obtained  for   $T \simeq 131$ MeV, $\mu_B \simeq 539$ MeV,  $\mu_{I3} \simeq -16$ MeV (left panel), whereas  for  $\sqrt{s_{NN}} = 17 $ GeV (right panel) it is obtained for    $T \simeq 147.6$ MeV, $\mu_B \simeq 218$ MeV,  $\mu_{I3} \simeq -2.1$ MeV.  
A yield ratio of two particles is denoted by the ratio of their respective symbols.
}
  \label{ratios_sNN}
\end{figure}

As one can see from the left panel of  Fig. \ref{ratios_sNN} the  quality of the fit achieved for $\sqrt{s_{NN}} = 4.9$ GeV is extremely good even for ${\bar \Lambda}/{\Lambda}$ and $ \Lambda/\pi^-$ ratios,  i.e. for the most problematic ratios  of \cite{KABAndronic:05}. 
 This is related to an essential improvement  of  the kaons and their ratios in the present model. 
 Thus, the ${\bar \Lambda}$ anomaly \cite{KABAndronic:05,AGS_L3} is not seen at this energy.

In the SPS energy range we used only the NA49 mid-rapidity data  for all ratios. 
There are two main reasons for such a selection. First, the NA49 are self-consistent and have relatively  small error bars for all energies.   Second, it is well known that right these  data are traditionally the most difficult ones  to be described  within the thermal model  \cite{KABAndronic:05, KABTiwari:12,KABStrangeness_horn_sigma_meson, KABthemal:2,  KABthemal:3}. 
Therefore, in order to demonstrate the new possibilities of the multi-component hadron  resonance model  we concentrate on the  NA49 data  fitting. 
In contrast to \cite{KABAndronic:05},  we included into the fit procedure  $\Omega/\pi^-$ and 
$\Xi/\pi^-$ ratios, but excluded from it  the dependent   $\Xi/\Lambda$ and $\Omega/\Xi$ ratios for hyperons. 
The results for the highest SPS energy $\sqrt{s_{NN}}= 17.3 $ GeV are  shown in the right panel 
of  Fig.  \ref{ratios_sNN}. 
These results are compared to the NA49 mid-rapidity  data for pions, kaons and (anti)protons \cite{KABNA49:17a,KABNA49:17b}, for the  set of  strange (anti)hyperons \cite{KABNA49:17Ha,KABNA49:17Hb,KABNA49:17Hc} and for $\phi$ meson 
\cite{KABNA49:17phi}. 

The variation of  the $R_\pi$ and $R_K$ radii  immediately allowed us to notably improve the description of $K^+/\pi^+$ ratio and all the ratios involving the  strange hyperons and pions (for instance, look at  $\Xi^-/\pi^-$ and $\Omega/\pi^-$). This also led  to a slight  improvement of   $K^-/K^+$ ratio. 
However, a slight change for the total strangeness of kaons means a larger change of the strange hyperons densities.  Although the most problematic ratios at this energy, namely $\bar \Lambda/\Lambda$,  $\bar \Xi^-/\Xi^-$ are 
 $\bar \Omega/\Omega$,  are improved only marginally compared to \cite{KABAndronic:05}, but as one can see from the right panel of   Fig.  \ref{ratios_sNN},  the crossed ratios of $\Xi/\Lambda$ and $\Omega/\Xi$,  which were not fitted, are  automatically reproduced well.
The obtained results for the chemical  FO temperature $T$ and baryonic chemical potential  $\mu_B$ almost coincide with the values $T \simeq 152$ MeV, $\mu_B \simeq 226$ MeV found  in \cite{KABAndronic:05} 
for this energy for the fitting the NA49 data alone. However, the resulting quality of our fit at this energy of collision  is essentially better: $\chi^2/dof \simeq 1.57 $ determined  in this work against $\chi^2/dof \simeq  2.78$  found  in \cite{KABAndronic:05}.

The same trend is seen for all  the SPS energies: a small variation of  kaon hard-core radius and 
a vanishing  pionic hard-core radius systematically improve the fit quality of the NA49 data. 
After such a  fitting of the most `hard' data it is clear  that a high quality description  of other data sets (or of all data sets)  is possible, if the experimental  data of different collaborations are reanalyzed and become consistent with each other. 

Since  the RHIC high energy  data of different collaborations agree with each other, we just analyzed  the STAR results  for   $\sqrt{s_{NN}}= 130$ GeV \cite{KABstar:130a,KABstar:130b,KABstar:130c,KABstar:200a} and $ 200 $ GeV \cite{KABstar:200a,KABstar:200b,KABstar:200c}.  
For the main subject of the present work the exotic ratios are not of a great importance and, hence, 
for the RHIC energies we fitted the same set of hadronic ratios  as for the highest SPS energy 
(see the right panel of Fig. \ref{ratios_sNN}). 
The variation of the pionic and kaonic hard-core radii practically does not affects the  fit quality 
at the RHIC energies. This is clearly seen from the comparison of the best fit results  for   $\sqrt{s_{NN}}= 130$ GeV found here $T \simeq 163.1$ MeV, $\mu_B \simeq 27.3$ MeV and the values  $T \simeq 162.5 \pm 5.5$ MeV, $\mu_B \simeq 35 \pm 11$ MeV found  in \cite{KABAndronic:05}  for a combined fit of 
PHENIX and STAR data assuming that the pion data  do not contain any contribution 
from weak decays. The better agreement is seen  for  the chemical FO parameters at $\sqrt{s_{NN}}= 200$ GeV obtained here  $T \simeq 162.2$ MeV, $\mu_B \simeq 17.6$ MeV and the values 
$T \simeq 160.5 \pm 2$ MeV, $\mu_B \simeq 20 \pm 4$ MeV found in 
\cite{KABAndronic:05}  for a combined fit of 
PHENIX and STAR  data excluding  $\bar p/\pi^-$ and $\phi/K^-$ ratios.

\begin{figure}[htbp]
\centerline{  \includegraphics[height=7 cm]{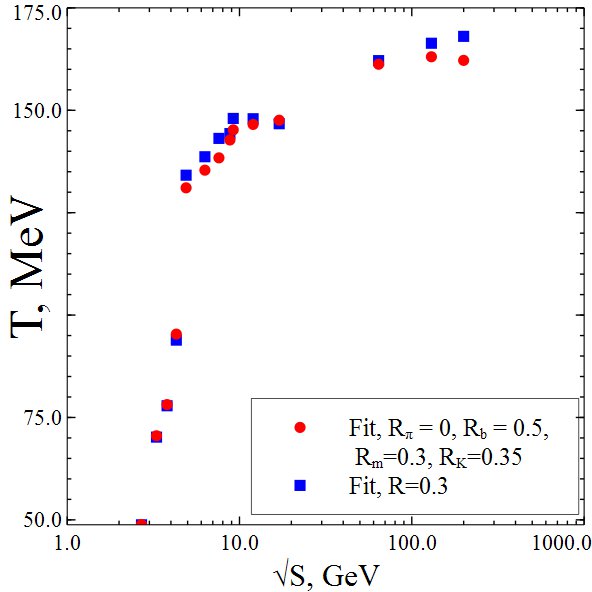} 
\hspace*{.5cm}
\includegraphics[height=7 cm]{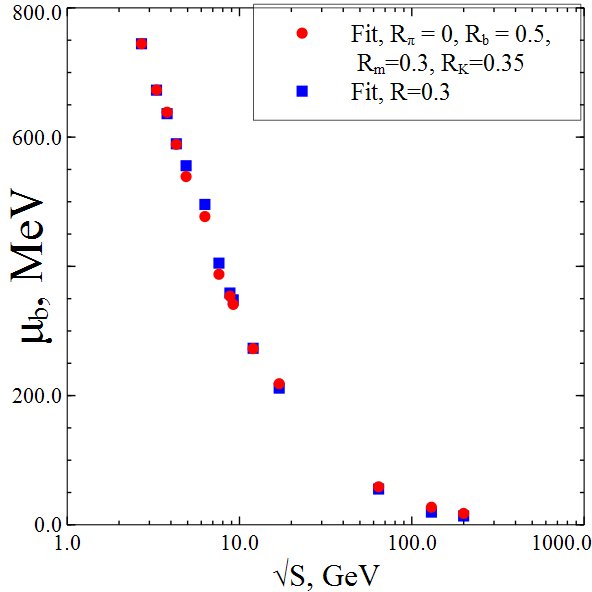} }  
 \caption{$\sqrt{s_{NN}}$  dependence of the chemical FO temperature (left panel) and  baryonic chemical potential (right panel)
 found here  within  the multi-component model (circles). For a comparison the corresponding quantities for a model with a single hard-core radius $R=0.3$ fm
 \cite{KABOliinychenko:12}
 (squares) are also shown.}
  \label{T_mu_R}
\end{figure}

\begin{figure}[htbp]
\centerline{    \includegraphics[height=7 cm]{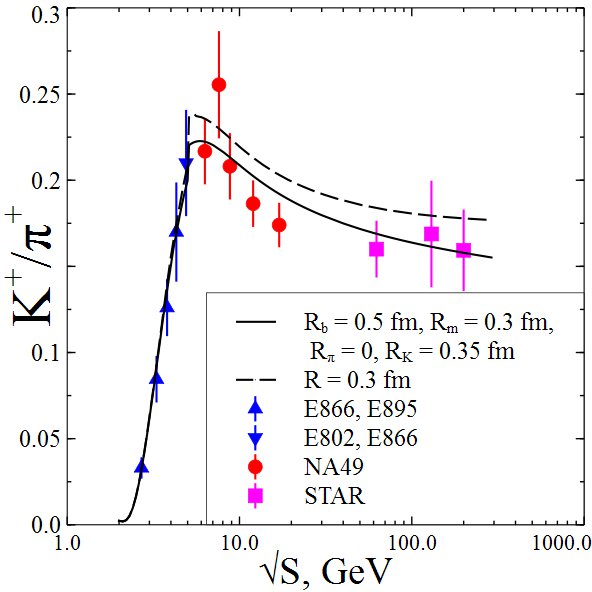}   
\hspace*{1.cm}
\includegraphics[height=7.cm]{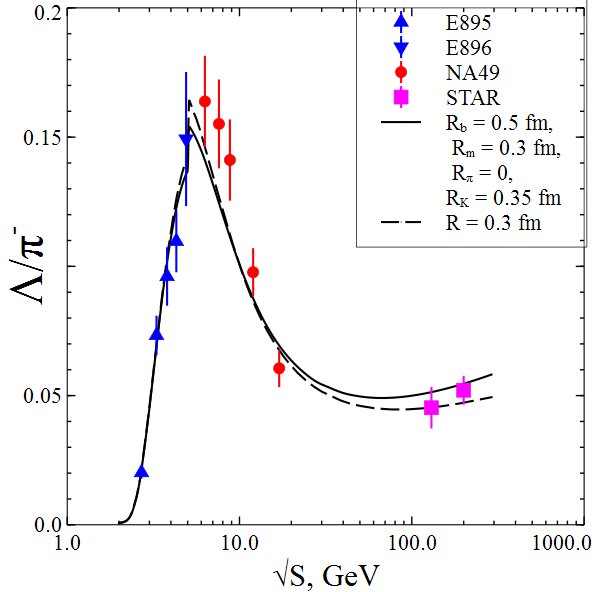} 
}
 \caption{$\sqrt{s_{NN}}$  dependences  of  $K^+/\pi^+$  (left panel)
 and  $\Lambda/\pi^-$ (right panel) ratios 
  obtained  here  within  the multi-component model are compared to that ones 
  found within the  one component model  \cite{KABOliinychenko:12}.
  }
  \label{horn_s}
\end{figure}

The above results are not surprising, since 
from Fig.  \ref{T_mu_R} it is  seen that the present multi-component fit almost reproduces  the 
values of the chemical FO temperature and baryonic chemical potential obtained in the model with 
a single hard-core radius for all hadrons  \cite{KABOliinychenko:12} which in its turn very well reproduces the results of  \cite{KABAndronic:05} for hadronic yield  ratios. 
However, the main result obtained within the present model, the best description of the Strangeness Horn, i.e. $K^+/\pi^+$ ratio,  alone with  $\Lambda/\pi^-$ ratio, is  demonstrated  in Fig. \ref{horn_s}.  These two ratios were thoroughly studied within the one component hadron resonance gas model  in \cite{KABAndronic:05} and then their analysis (together with other ratios) was continued 
in  \cite{KABStrangeness_horn_sigma_meson}. As we mentioned earlier  for the one component model the fit of ratios is not affected by the excluded volume correction. If, however, one imposes some additional constraints like the artificial baryon charge conservation criticized in \cite{KABOliinychenko:12}, then some small correction (below 5 \%) can appear.  This is the reason why the one component model  fit of  $K^+/\pi^+$ and  $\Lambda/\pi^-$ ratios found in \cite{KABAndronic:05, KABStrangeness_horn_sigma_meson} for  a single hard-core radius  $R=0.3$ fm is essentially worse than  the one component model fit with the same hard-core radius shown in Fig.  \ref{T_mu_R}.  Evidently, in Refs. \cite{KABAndronic:05, KABStrangeness_horn_sigma_meson} the fit of 
$\Lambda/\pi^-$ ratio  indicates a problem with too steep rise 
as a function of  $\sqrt{s_{NN}}$ compared to the data, while both of these ratios show too slow decrease compared to the data. Note that such a behavior of  $K^+/\pi^+$ and  $\Lambda/\pi^-$  ratios is typical for almost all statistical models (see Figs. 7 and 8 and the corresponding discussion  in \cite{KABTiwari:12}).
Evidently, the too steep rise in  $\Lambda/\pi^-$  behavior is   a consequence of  the  ${\bar \Lambda}$ anomaly \cite{KABAndronic:05,AGS_L3} discussed above.  The one component fit of the present approach does not indicate such difficulties for  $\Lambda/\pi^-$ ratio, while the slow decrease in  $K^+/\pi^+$  ratio still is there. 
However, the multi-component approach  really removes such a defect in $K^+/\pi^+$  without spoiling the other ratios including $\Lambda/\pi^-$ one as it is seen from Fig.  \ref{T_mu_R}. The results for other ratios are available and will be shown in a longer work. 

Actually the best fit for $K^+/\pi^+$  ratio found here practically coincides with the dashed curve drawn 
in Fig. 4 of   \cite{KABStrangeness_horn_sigma_meson} which was obtained assuming  an 
existence of the Hagedorn mass spectrum  of  hadrons \cite{KABlimitT}. However, we do not share
this  hope of the authors of   Ref. \cite{KABStrangeness_horn_sigma_meson}. Also  we do not agree with  such an estimate
and, hence, we cannot accept it  as a  real solution of the puzzling problem. Our skepticism  is based on the following facts. First of all, we note that inclusion of the hypothetical states  with the masses up to 3 GeV should essentially modify not only $K^+/\pi^+$ results, but all other ratios in uncontrollable way. Then the authors 
of   \cite{KABStrangeness_horn_sigma_meson} would spoil their own results reported in many nice works. 
Second, it is unclear to us why in this case one should stop at hadron mass of 3 GeV and not to increase it till 10 GeV or even to 100 GeV? It is clear that already  in the former case the one component excluded volume description would lead to a whole complex of problems that are typical to the Hagedorn mass spectrum and 
in order to get rid of them one  will   have to unavoidably introduce the excluded volume which is proportional to a mass of a heavy resonance. Such models are well known  \cite{FWM:08, FWM:09}, but then in the framework of such models the heavy  hadronic resonances should be regarded as quark-gluon bags and the whole treatment should be completely changed.

Furthermore,  in \cite{KABStrangeness_horn_sigma_meson} the estimates based on  the Hagedorn mass spectrum inclusion   were  done without accounting for the large resonance width and without knowing the branching ratios of these hypothetical resonances.
Although the  mass dependence of width of heavy resonances was found within the finite width model of quark gluon bags \cite{FWM:08, FWM:09} and recently  it was successfully verified on the Regge trajectories of heavy mesons \cite{KABRegge:11}, the possible channels of their decays and  the corresponding 
branching ratios   are completely unknown yet. However, the more serious issue is that the finite width model 
\cite{FWM:08, FWM:09}  explains that the huge deficit of the empirical hadronic spectrum compared to the Hagedorn one is due to the fact that the `missing' hadrons with the masses above  2.5 GeV and with  the  large width  
  are the quark gluon bags, which  are extremely suppressed (by fifteen-sixteen orders of magnitude) compared to the  stable hadrons up to the temperatures of about half of the Hagedorn temperature. Therefore, we again come to a conclusion that the `missing' hadrons should not be included into 
the hadron resonance gas model spectrum, but they should  be attributed to the  spectrum of quark gluon bags. The practical consequence out  of  these facts is as follows: due to the short life time $\tau \simeq  \sqrt{\frac{M_0}{M}}0.5$ fm/c of the bag of mass $M \ge M_0 \simeq 2.5$ GeV, by the time of chemical FO such bags should have been, probably, completely  decayed into the stable hadrons and light hadron resonances. Since up to the  moment  of chemical FO the chemical equilibrium is assumed to exist,
then all thermodynamic quantities of pions (or other particles appeared from these bags) should (locally)  have their equilibrium values in accordance with the hadron resonance gas model spectrum, i.e. at chemical FO the result of the bag decays should not bee seen.

\section{Conclusions}

In the present work we considered the hadron resonance gas model with multi-component hard-core radii, i.e. we treat the pion $R_\pi$ and  kaon $R_K$ hard-core radii as independent fitting parameters compared to the  hard-core radius of baryons $R_b=0.5$ fm and that one of all other mesons $R_m = 0.3$ fm.   Such an approach allows us to essentially improve the quality of the global fit of  hadron yield ratios derived from  the mid-rapidity data measured at the AGS, SPS and two highest RHIC energies. Thus, for $R_\pi = 0$ fm  and  $R_K=0.35$ fm we found  $\chi^2/dof \simeq 1.018$ which is  the best value for the global fit  compared to other analyses. It is necessary to stress 
that at SPS energies the present fit included only the NA49 data, which are usually  hard to be fitted by the hadron resonance gas model. The suggested  approach allows us to drastically improve the quality of the data description and, as a consequence,  it is  able  to completely describe the Strangeness Horn irregular behavior for the first time.  Thus, the developed approach  gives  a simple solution to the  puzzle of the Strangeness Horn   description   without the need to use   the hypothetical hadron resonances of masses up to 3 GeV which, so far,  are not observed in the experiments.

The found small values of hard-core radii are consistent with the results of  analysis done in 
\cite{KABRaju:92} that up to the temperatures of  about 170 MeV the hadron-hadron repulsive and attractive interaction contributions into the system pressure  practically compensate each other. In fact, we determined the second virial coefficients 
of hadrons using the statistical model with the multi-component hard-core  repulsion. Evidently, such an approach can be used  to further improve the description of other particle ratios. Thus,   the suggested multi-component model  provides us with a practical  way   to extract the second virial coefficients for all hadrons and tabulate them as a  function of temperature as this  is done  for  usual gases. The main problem, however, is related to the poor quality of  existing experimental data. The present analysis clearly shows that to accurately determine the hadronic second virial coefficients and, thus, to provide the community with the data allowing in principle to extract the statistical measure of interaction for any pair of hadrons we need much better data up to   $\sqrt{s_{NN}} \simeq 20 $ GeV.  We hope that 
the  Dubna Nuclotron and  the future colliders NICA and FAIR will successfully  resolve at least the half of  this task. 

\vskip3mm

{\bf Acknowledgments.} We would like to thank A. Andronic for  providing an access to well-structured experimental data. 
The authors are thankful to M. Gazdzicki for valuable comments. K.A.B. and G.M.Z. acknowledge the partial  support of the Program 'Fundamental Properties of 
Physical Systems under Extreme Conditions' launched by the Section of Physics and Astronomy of National Academy of  Sciences of Ukraine.
The work of  A.S.S.  was supported in part by the Russian
Foundation for Basic Research, Grant No. 11-02-01538-a.



\begin{thebibliography}{99}


\bibitem{KABAndronic:05}
%
A. Andronic, P. Braun-Munzinger and J. Stachel,  
Nucl. Phys. A {\bf 772}, 167 (2006) and references therein.

\bibitem{KABAndronic:12}
A. Andronic, P. Braun-Munzinger, J. Stachel and M. Winn,
arXiv:1201.0693 [nucl-th].

\bibitem{KABbig_radii}
G.D. Yen, M.I. Gorenstein, W. Greiner and S.N. Yang,
 Phys. Rev. C {\bf 56}, 2210 (1997).
 
 \bibitem{KAB_two_comp_VdW}
%
G. Zeeb, K.A. Bugaev, P.T. Reuter and H. St\"ocker,
Ukr. J. Phys.  {\bf 53},  279 (2008).

\bibitem{KABOliinychenko:12}
%
D.R. Oliinychenko, K.A. Bugaev and  A.S. Sorin, 
arXiv:1204.0103 [hep-ph].  

\bibitem{KABCleymansFO}
%
%
J. Cleymans, K. Redlich, Phys. Rev. Lett. {\bf 81}, 5284 (1998).

\bibitem{KABTiwari:12}
%
S. K. Tiwari,  P. K. Srivastava, and C. P. Singh,
Phys. Rev.  C {\bf 85}, 014908 (2012).  

\bibitem{KABStrangeness_horn_sigma_meson}
A. Andronic, P. Braun-Munzinger and J. Stachel,
Phys. Lett. B {\bf 673}, 142 (2009).

\bibitem{Gazd_Horn:99}
%
M. Gazdzicki and M. I. Gorenstein, Acta Phys. Polon. B {\bf  30}, 2705 (1999).

\bibitem{Step}
%
M. I. Gorenstein, M. Gazdzicki and K. A. Bugaev,
Phys. Lett. B {\bf 567}, 175  (2003).

\bibitem{Gazd_rev:10}
%
M. Gazdzicki, M. I. Gorenstein and P. Seyboth,
Acta Phys. Polon. B {\bf 42}, 307 (2011).

\bibitem{KABRaju:92}
%
R. Venugopalan and M. Prakash, Nucl. Phys. A {\bf 546}, 718 (1992). 

\bibitem{KABPBM:99}
%
P. Braun-Munzinger, I. Heppe and J. Stachel, Phys. Lett. B{\bf  465},  15 (1999). 

\bibitem{KABKostyuk:99}
%
M. I. Gorenstein, A. P. Kostyuk and Y. D. Krivenko,  J. Phys. 
G {\bf 25},  L75 (1999). 

\bibitem{Bugaev_Lorentz_cont_2}
K.A. Bugaev,
Nucl. Phys. A {\bf 807},   251 (2008);
arXiv:1012.3400 [nucl-th].

\bibitem{KABFisher_67}
M. E. Fisher, Physics {\bf 3}, 255 (1967).

\bibitem{KABDillmann:91} 
%
A. Dillmann and G. E. Meier, 
J. Chem. Phys. {\bf 94}, 3872 (1991). 

\bibitem{KABLFK:94} 
%
A. Laaksonen, I. J. Ford, and M. Kulmala,
 Phys. Rev. E {\bf 49}, 5517 (1994).
 
\bibitem{KABBondorf}
%
J. P. Bondorf et al., 
 Phys. Rep. {\bf 257}, 131 (1995)
and references therein. 

\bibitem{KABBugaev:00}
K. A. Bugaev,
M. I. Gorenstein, I. N. Mishustin and W. Greiner,
Phys. Rev. {\bf C62},  044320 (2000);  Phys. Lett. {\bf B 498},  144 (2001) and references therein.

\bibitem{THERMUS}
S. Wheaton and J. Cleymans,
arXiv:0407174 [hep-ph].

\bibitem{AGS_pi1}
%
J.L. Klay et al. (E895), Phys. Rev. C {\bf 68}, 054905 (2003). 

\bibitem{AGS_pi2}
%
 L. Ahle et al. (E866/E917), Phys. Lett. B {\bf 476},  1 (2000); Phys. Lett. B {\bf 490}, 
 53 (2000).

\bibitem{AGS_p1}
%
 B.B. Back et al. (E917), Phys.  Rev. Lett. {\bf 86}, 1970  (2001).

\bibitem{AGS_p2}
%
J.L. Klay et al. (E895), Phys. Rev. Lett. {\bf 88},  102301 (2002). 

\bibitem{AGS_L}
%
C. Pinkenburg et al. (E895), Nucl. Phys. A {\bf 698},   495c (2002).


\bibitem{AGS_Kas}
%
P. Chung et al. (E895), Phys. Rev. Lett. {\bf 91}, 202301  (2003).


\bibitem{AGS_phi}
%
 B.B. Back et al. (E917), Phys. Rev. C {\bf 69},   054901 (2004).


\bibitem{AGS_L2}
%
S. Albergo et al. (E896),  Phys. Rev. Lett. {\bf  88},   062301 (2002).

\bibitem{AGS_L3}
%
B.B. Back et al. (E917), Phys. Rev. Lett. {\bf 87},   242301  (2001).  


\bibitem{KABthemal:2}
%
F. Becattini, M. Gazdzicki, A. Keranen, J. Manninen, R. Stock, Phys. Rev. C {\bf  69},  
024905  (2004).

\bibitem{KABthemal:3}
%
A. Dumitru, L. Portugal, D. Zschiesche, Phys. Rev. C {\bf 73}, 024902 (2006).

\bibitem{KABNA49:17a}
%
 S.V. Afanasiev et al. (NA49), Phys. Rev. C {\bf 66},  054902 (2002).
 
\bibitem{KABNA49:17b}
%
 S.V. Afanasiev et al. (NA49), Phys. Rev. C {\bf 69},  024902 (2004).  
 
%
 

\bibitem{KABNA49:17Ha}
%
T. Anticic et al. (NA49), Phys. Rev. Lett. {\bf 93},   022302  (2004).

\bibitem{KABNA49:17Hb}
%
S.V. Afanasiev et al. (NA49), Phys. Lett. B  {\bf 538},   275 (2002).

\bibitem{KABNA49:17Hc}
%
C. Alt et al. (NA49), Phys. Rev. Lett. {\bf 94},   192301 (2005).

\bibitem{KABNA49:17phi}
%
S.V. Afanasiev et al. (NA49), Phys. Lett. B  {\bf 491},   59 (2000).


\bibitem{KABstar:130a}
%
J. Adams et al. (STAR), Phys. Rev. Lett. {\bf 92},   182301 (2004).

\bibitem{KABstar:130b}
%
 J. Adams et al. (STAR), Phys. Lett. B {\bf 567},  167 (2003).

\bibitem{KABstar:130c}
%
C. Adler et al. (STAR), Phys. Rev. C {\bf 65},  041901(R) (2002).

\bibitem{KABstar:200a}
%
J. Adams et al. (STAR), Phys. Rev. Lett. {\bf  92},  112301 (2004).

\bibitem{KABstar:200b}
%
 J. Adams et al. (STAR), Phys. Lett. B {\bf  612},  181  (2005).

\bibitem{KABstar:200c}
%
A. Billmeier et al. (STAR), J. Phys. G {\bf 30},   S363 (2004).



\bibitem{KABlimitT}
%
R. Hagedorn, Suppl. Nuovo Cimento {\bf 3}, 147 (1965).

\bibitem{FWM:08}
%
K. A. Bugaev, V. K. Petrov and G. M. Zinovjev, 
Europhys. Lett. {\bf 85},  22002 (2009) and references therein.

\bibitem{FWM:09}
%
K. A. Bugaev, V. K. Petrov and G. M. Zinovjev, 
Phys. Rev. C {\bf 79}, 054913  (2009) and references therein.

\bibitem{KABRegge:11}
%
K. A. Bugaev, E. G. Nikonov, A. S. Sorin and G. M. Zinovjev,
 JHEP 02,  059 (2011) . 






\end{thebibliography}
\end{document}